\documentclass[pra,twocolumn,amsmath,showpacs]{revtex4}

\usepackage{graphicx}
\usepackage{dcolumn}
\usepackage{bm}
\usepackage{color}
\usepackage{subfigure}
\usepackage{amsmath}
\def\beq{\begin{equation}} \def\eeq{\end{equation}}
\def\beqa{\begin{eqnarray}} \def\eeqa{\end{eqnarray}}
\def\bce{\begin{center}}  \def\ece{\end{center}}
\def\bfig{\begin{figure}}  \def\efig{\end{figure}}
\def\bpic{\begin{picture}}  \def\epic{\end{picture}}
\def\bit{\begin{itemize}}    \def\eit{\end{itemize}}
\def\ben{\begin{enumerate}}    \def\een{\end{enumerate}}
\def\l<{\left<}   \def\r>{\right>}

\begin{document}

\title{Collapse events of two-color optical beams.}

\author{Alexey Sukhinin$^{1}$, Alejandro B Aceves$^{2}$, Jean-Claude Diels$^{3}$, Ladan Arissian$^{3}$}

\affiliation{1. Department of Mathematics and Statistics, University of Vermont, Burlington, VT 05405, USA\\
2. Department of Mathematics, Southern Methodist University, Dallas, TX 75275, USA\\
3. Department of Electrical and Computer Engineering, University of New Mexico, Albuquerque, NM, 87131, USA\\
}


\date{\today}

\begin{abstract}
In this work, we study optical self-focusing that leads to collapse events for the time-independent model of co-propagating beams with different wavelengths. We show that
collapse events depend on the combined critical power of two beams for both fundamental, vortex and mixed configurations as well as on the ratio of their individual powers.
\end{abstract}

\pacs{42.65.Jx, 42.65.Sf, 42.65.Tg, 52.38.Hb, 52.35.Sb}

\maketitle
Gaussian self-focusing is one of the most important effects in nonlinear optics
and laser filamentation. It is well known that at high intensities,
laser beams tend to self-focus in the propagation through nonlinear media
with focusing third order nonlinearity (Kerr nonlinearity).
For quasi-monochromatic waves, the two dimensional Nonlinear Schr\"odinger
Equation (NLSE) is a precise mathematical model that describes such phenomenon.
In this case, rigorous results such as the virial theorem support the
fact that an initial beam having power
above critical (as defined by an exact balance between
self-focusing and diffraction)
 leads to a collapse event
at finite propagation distance~\cite{Chiao64}. As the beam propagates towards
the collapse point, it assumes a particular shape known as the Townes soliton
solution of the NLSE~\cite{Moll03}.
If instead, the power is below critical the beam diffracts.
As the beam approaches the collapse point, with the increasing intensity on axis,
higher order nonlinear effects than Kerr come into effect. Typically, these
higher order nonlinearities are de-focusing and collapse may be arrested such that a ''filament'' is formed~\cite{Akhmanov66,Chalus08}.
The phenomenon of filamentation has gained considerable interest
with the discovery of filaments in air in the near IR~\cite{Braun95}
and in the UV~\cite{Zhao95}, as a way to remotely direct light at high intensities.
Unfortunately, single filaments were only observed to exist over
distance of the order of meter, rather than the anticipated km.
In this paper  we enrich the model by considering the
role of orbital angular momentum on the beam (vortex) propagation and doing this for the co-propagation of light in pairs of wavelengths.  Recently, vortex beams were generated
at 800 nm and used for the control of filamentation~\cite{Fisher06} and were investigated theoretically in the context of
UV filaments in air~\cite{Sukhinin13, Chaitanya}. Vortex filaments can generate conical high-harmonic generation~\cite{lin} and in principle filaments with angular momentum  can be an effective source of high harmonics~\cite{Gariepy}. Co-propagating pairs of filaments have important applications in remote impulsive stimulated Raman scattering, where the IR fs pulse ''impulsively'' excites a vibration or rotation, stimulating the Raman emission pumped by the picosecond UV filament~\cite{Feng15c}. Nonlinear interactions of multi-color laser beams in plasmas have been studied theoretically~\cite{Yi09}. Recent  work~\cite{Sukhinin15, Doussot16} suggests the co-existence of two-color optical filaments at resonant frequencies ({i.e. frequencies that are multiple of each others}).

The propagation  of combined
beams having different wavelengths and possibly propagating
with different spatial profiles is considered here.
In particular conditions leading to collapse events
with two-color, time-independent modes
 assuming to be not at resonance
are being derived.
 The model studied can be viewed as a generalization of the two
dimensional Nonlinear Schrodinger Equation. We first find stationary solutions and then numerically study their behavior on propagation. Our main outcomes
are first that the co-propagating filaments collapse
simultaneously when the sum of their powers is above
a combined critical power with power ratio close to the stationary states. Perhaps more surprisingly,
the same is true for two-color fundamental-vortex and
vortex-vortex configurations. It is the particular
evolution towards the collapse event that differs as
we will show its dependence on the initial state.
Altogether the stationary solutions, in the absence of higher
order nonlinear terms~\cite{Chalus08} are always unstable
as it is the case for 2D NLSE~\cite{Weinstein85}.

Consider the system of two time-independent non-dimensional
equations describing the co-propagation of optical beams in a
non-resonant regime, coupled by combined Kerr effect
responsible for self-focusing

\begin{eqnarray}
\label{main1}
\frac{i}{\gamma}\frac{\partial E_{1}}{\partial z} +  \Delta_{\bot}E_{1} + \frac{1}{\gamma^2}(|E_{1}|^2 + 2|E_{2}|^2)E_{1}  = 0 \\
\label{main2}
i\frac{\partial E_{2}}{\partial z} + \Delta_{\bot}E_{2} +(2|E_{1}|^2 + |E_{2}|^2)E_{2} = 0
\end{eqnarray}
where $E_i=E_i(x,y,z)$, $\Delta_{\bot}=\frac{\partial^2 }{\partial x^2}+\frac{\partial^2}{\partial y^2}$ and $\gamma$ is non-resonant coefficient, $\omega_2=\gamma\omega_1$.

We begin our analysis by summarizing  known properties of the Townes soliton solution of the two dimensional Nonlinear Schrodinger Equation
\[i\frac{\partial E}{\partial z} +  \Delta_{\bot}E + |E|^2E=0\]

For radially symmetric profiles, the field $E$ takes the form $E(x,y,z)=F(r)e^{i\beta z}$, where $r$ is the radius, $\beta$ propagation constant and $F$  is the mode amplitude. The stationary profile is then obtained by solving the nonlinear eigenvalue problem

\begin{eqnarray}
\label{anzatz1}
\frac{1}{r}\frac{dF}{dr}+ \frac{d^2F}{dr^2} + F^3=\beta F \end{eqnarray}
for which the ground state is know as the Townes soliton.
The existence and properties of the soliton
has been studied in great detail{~\cite{Chiao64}. 
 
In fact, it is a family of eigensolutions
having the same critical power but different  amplitude/width values.
This is due to the fact that equation (3) has the invariant scaling
property $F(r) = \sqrt{\beta}f(s)$, $s= \sqrt{\beta} r$
which makes the equation parameter free.
In physical variables, the critical power for self focusing is given
by $P_{cr}= \alpha \lambda^2/(4 \pi n_0 n_2)$, where $n_0$ and $n_2$ are
the linear and non-linear indexes of refraction, $\lambda$ the wavelength
and the power of the Townes soliton is equal to $\alpha$.

Self-focusing leading to collapse occurs if the initial power
of the beam is greater than the critical power. It was later
shown that collapsing beam eventually converges to the
Townes profile for $\beta\rightarrow \infty$. There is no explicit formula for Townes solution, however,
the nonlinear eigenvalue problem can be solved numerically
using Newton's method. The value of $\alpha$ is estimated
to be 
\[\alpha =\int |F(r)|^2rdr \approx 1.86225\]
We now take the same approach to solve equations (1)-(2), namely we seek radially symmetric stationary solutions of the  form
\begin{eqnarray}
\label{anzatz1} \nonumber
E_{1}(x,y,z) = {\cal E}_{1}(r)e^{i\beta_{1} z + im_{1}\theta}
\end{eqnarray}
\begin{eqnarray}
\label{anzatz2} \nonumber
E_{2}(x,y,z) = {\cal E}_{2}(r)e^{i\beta_{2} z + im_{2}\theta}
\end{eqnarray}
where now $\beta_{2}=\gamma\beta_{1}=\beta$ are the corresponding propagation constants and the integers $m_i$ correspond to vortex topological charges. If $m_1=m_2=0$, the following system of eigenvalue equations is obtained
\begin{eqnarray}
\label{building_block1}
\frac{1}{\gamma^2}\beta {\cal E}_{1} = \frac{1}{r}\frac{d{\cal E}_{1}}{dr} + \frac{d^2 {\cal E}_{1}}{d^2r} + \frac{1}{\gamma^2}({\cal E}_{1}^2 + 2{\cal E}_{2}^2){\cal E}_{1}\,\,\\
\label{building_block2}
\beta {\cal E}_{2} = \frac{1}{r}\frac{d{\cal E}_{2}}{dr} + \frac{d^2 {\cal E}_{2}}{d^2r} +(2{\cal E}_{1}^2 + {\cal E}_{2}^2){\cal E}_{2} \,\,
\end{eqnarray}

We observe that this system has the same invariant scaling property as in equation (3). We then make the substitution ${\cal E}_i(r) = \sqrt{\beta}f_i(s)$, $s= \sqrt{\beta} r$, which eliminate the parameter $\beta$. The significance of this property is that in the numerical search of radially symmetric states, we only need to find one solution of the family.

\begin{figure}[htbp]
  \begin{center}
    \mbox{
      \subfigure[]{\scalebox{0.5}{\includegraphics[width=3in,height=2in]{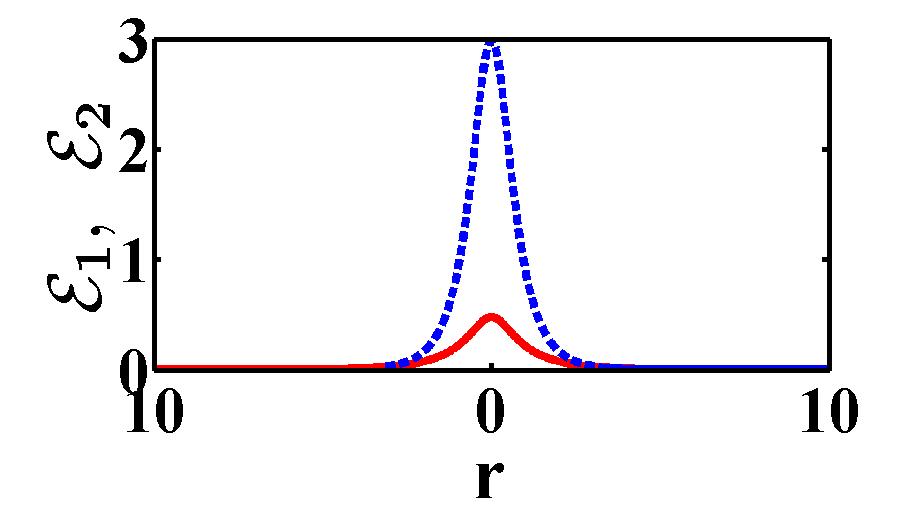}}}} \quad
      \subfigure[]{\scalebox{0.5}
{\includegraphics[width=3in,height=2in]{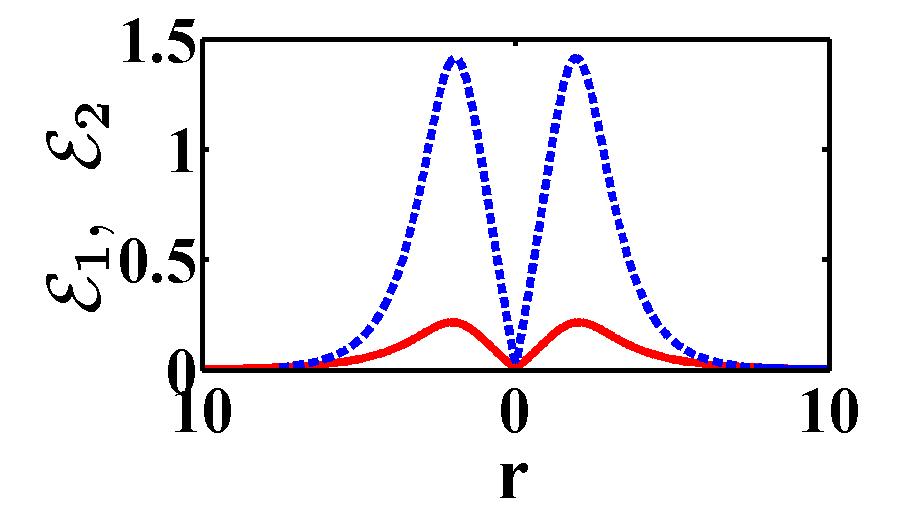}}}
      \subfigure[]{\scalebox{0.7}
{\includegraphics[width=3in,height=2in]{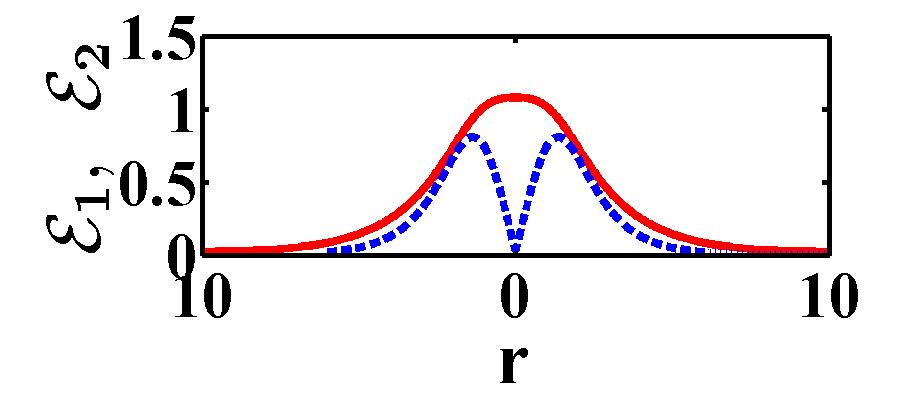}}}
      \subfigure[]{\scalebox{0.5}
{\includegraphics[width=3in,height=2in]{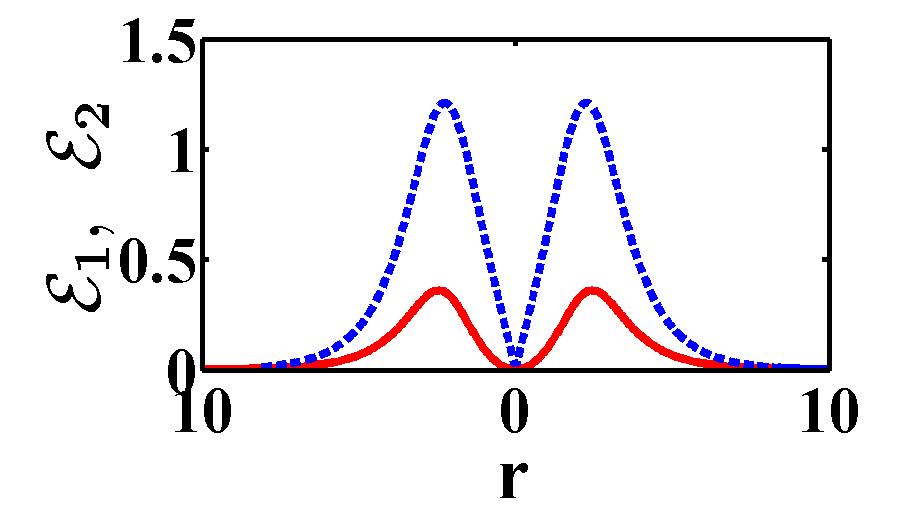}}}
      \subfigure[]{\scalebox{0.5}
{\includegraphics[width=3in,height=2in]{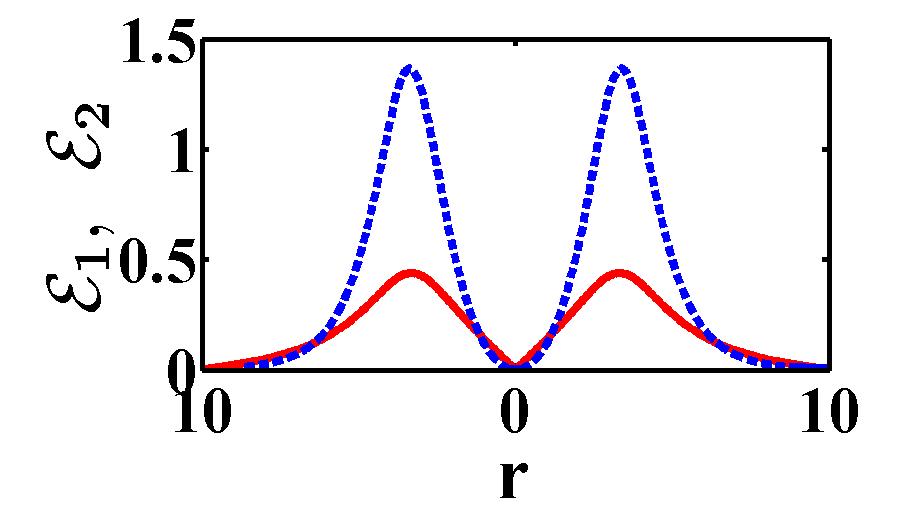}}}
    \caption{Profiles of the co-existing states of ${\cal E}_1$(solid red line) and ${\cal E}_2$(dashed blue line): (a) $m_1=m_2=0, \beta=2, \gamma=1.8, P_1=0.0846, P_2=1.773$, (b) $m_1=m_2=1, \beta=0.7, \gamma=1.75, P_1=0.2374, P_2=7.438$, (c) $m_1=0, m_2=1, \beta=0.7, \gamma=1.8, P_1=3.172
, P_2=1.7919$, (d) $m_1=2, m_2=1, \beta=0.6, \gamma=1.15, P_1=0.6256, P_2=6.9796$, (e) $m_1=1, m_2=2, \beta=0.8, \gamma=2.2, P_1=1.9479, P_2=11.94562$}
    \label{fig:m_0}
  \end{center}
\end{figure}

This system of equations has been solved numerically using Newton's method. Figure 1(a) shows the co-existing fundamental states having topological charge $m_i=0$, $i=1,2$ for eigenvalue $\beta=2$. While the total power of the  states ($P_{tot}=P_1+P_2$) depends on $\gamma$, it does not  depend on $\beta$. On Figure (1a), the total power $P_{tot} \approx 1.85771$ for $\gamma=1.8$, with  $P_1 \approx 0.0846148$ and $P_2 \approx 1.773$. Notice that both powers are smaller than that of the Townes soliton. Figure 1(b, d,e) represents vortex states with charges (1,1), (2,1) and (1,2) respectively. Again, within each family identified by the $(m_1,m_2)$ charges, $P_1$ and $P_2$ remain independent of $\beta$. Figure 1(c) represents a mixed state having vortex ${\cal E}_2$ with charge 1 "covered" by the fundamental state ${\cal E}_1$.

To determine propagation properties of these modes, what follows are numerical experiments where an incident field at $z=0$ is a  perturbed version of these states. We consider initial conditions where the total power is either above or below that of the stationary mode. In doing this, we are able to determine conditions leading to collapse by controlling the ratio of individual powers $P_1/P_2$. We do this by use of  direct numerical simulations of equations (1,2). The numerical scheme used was split-step Fourier scheme with up to $512 \times 512$ points in spatial discretization. 

\begin{figure}[htbp]
  \begin{center}
    \mbox{
      \subfigure[]{\scalebox{0.5}{\includegraphics[width=3in,height=2in]{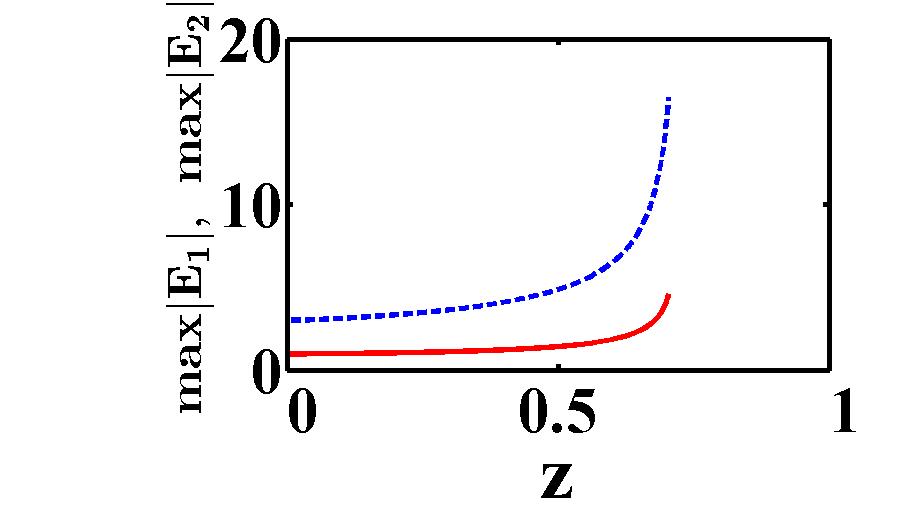}}}} \quad
      \subfigure[]{\scalebox{0.5}
{\includegraphics[width=3in,height=2in]{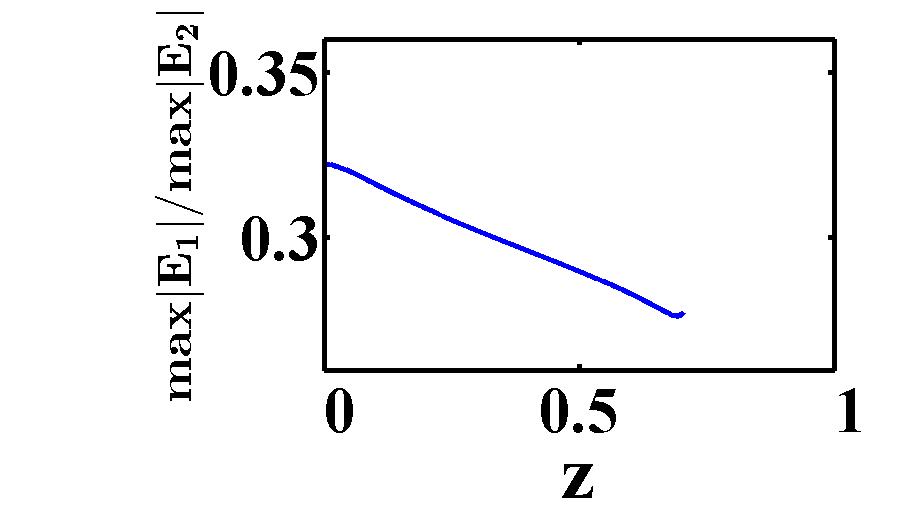}}}
      \subfigure[]{\scalebox{0.5}
{\includegraphics[width=3in,height=2in]{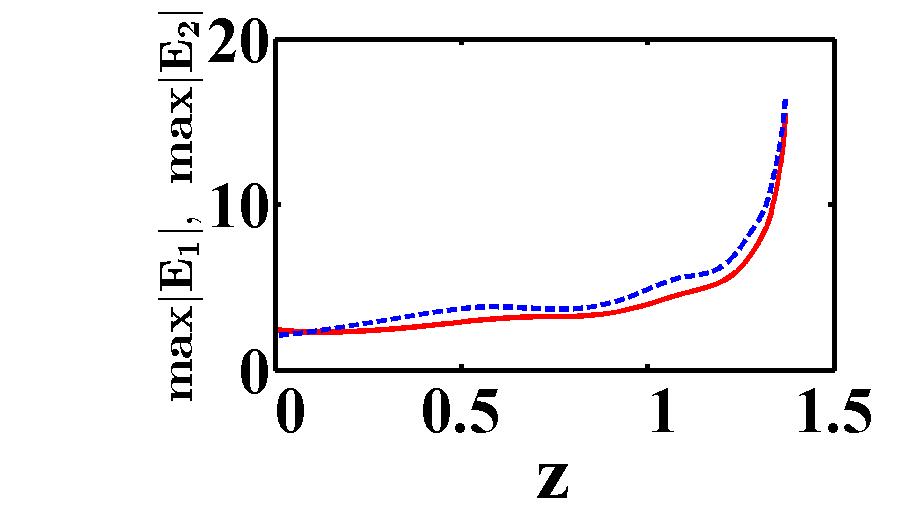}}} 
	  \subfigure[]{\scalebox{0.5}
{\includegraphics[width=3in,height=2in]{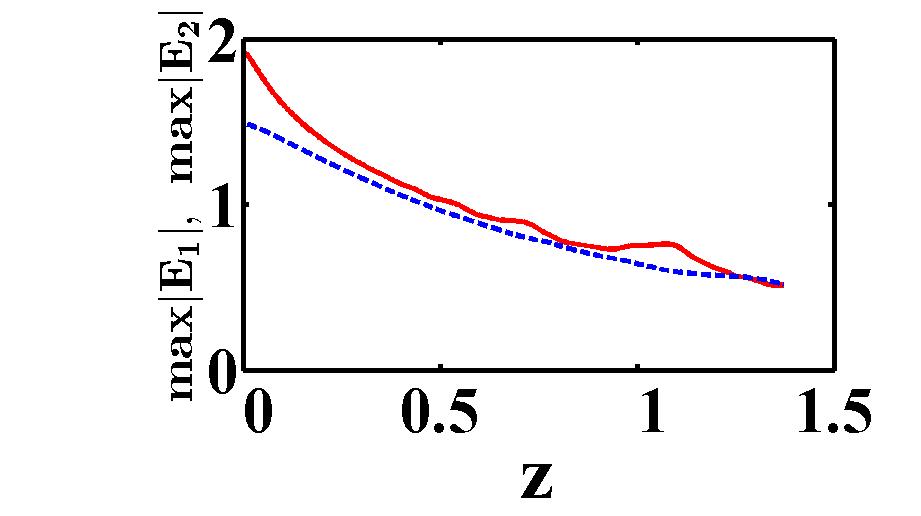}}} \quad
    \caption{$m_1=m_2=0$, $\gamma=1.8$. Simultaneous collapse: (a,b) initial conditions close to ground states $|E_1(z=0)|=2{\cal E}_{1},|E_2(z=0)|={\cal E}_{2}$, $P_{1_{in}}=4P_1, P_{2_{in}}=P_2,P_{tot_{in}=1.155P_{tot}}$, (c) initial conditions far from ground states $|E_1(z=0)|=5{\cal E}_{1},|E_2(z=0)|=0.7{\cal E}_{2}$, $P_{1in}=25P_1, P_{2in}=0.5P_2, P_{tot_{in}}=1.6P_{tot}$. Simultaneous diffraction: (d) initial conditions far from ground states with the total power above critical, $|E_1(z=0)|=5{\cal E}_{1},|E_2(z=0)|=0.5{\cal E}_{2}$, $P_{1_{in}}=25P_1,P_{2_{in}}=0.25P_2,P_{tot_{in}}=1.37P_{tot}$}
    \label{fig:m_0}
  \end{center}
\end{figure}

Figure 2(a) shows simultaneous self-focusing leading to collapse with the total power $P_1+P_2$ above that of the ground states. Instead, beams with power below critical diffract. This suggests that the power of the ground state is indeed a critical power condition as in the Townes soliton case. Collapse events are present for different perturbations of the initial beams power ratio. Furthermore, based on these simulations, we also conjecture that the collapsing beams with total power above critical which are close to the power ratio of the stationary modes evolve into the bound states of (4,5) corresponding to $\beta \rightarrow \infty$. This case highlights the fact that even if one of two beams has power less than its own "critical" value (${\cal E}_2$),  collapse is observed because the combined power  $P_1+P_2$ exceed the critical value.  However, if the beams ratio is far from that of the stationary mode, collapse is not guaranteed even if the total power still above critical. This scenario is shown on Figure 2(d). In this case both beams are far from their corresponding ground state equilibria. The initial power ratio also affects how the beams approach collapse. Based on our observations, collapse develops at different rates; in particular the amplitude of ${\cal E}_2$ increases faster. Having individual rates of collapsing beams could be important for modeling of the ionization processes.

For the next simulation we study the collapse conditions of a Gaussian beam combined with am $m=1$ vortex. The initial conditions are

\[{\cal E}_1(r)=A_1e^{-r^2/\omega_1^2}  \,\,\,\,\,\,\, {\cal E}_2(r)=A_2r e^{-r^2/\omega_2^2}\]

\begin{figure}[htbp]
  \begin{center}
    \mbox{
      \subfigure[]{\scalebox{0.5} {\includegraphics[width=3in,height=2in]{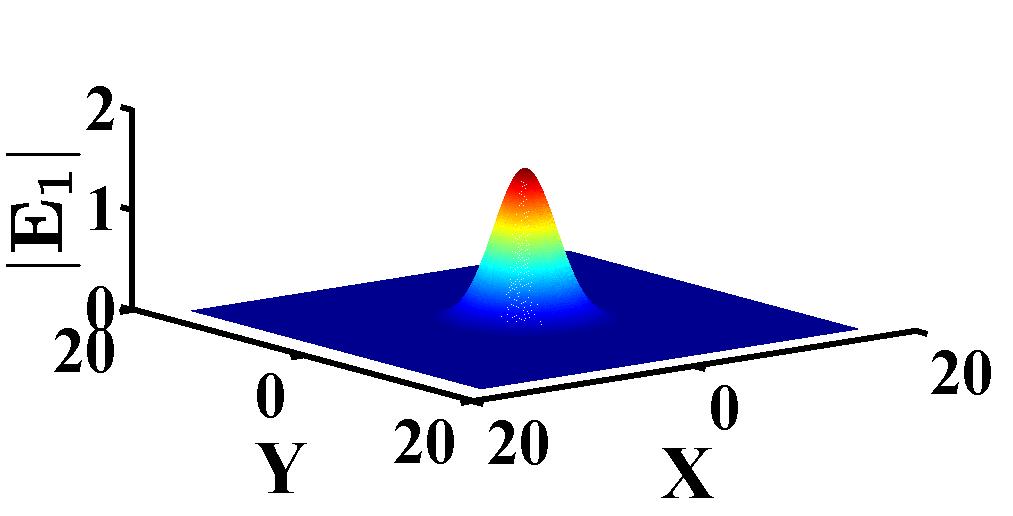}}}} \quad
      \subfigure[]{\scalebox{0.5}
{\includegraphics[width=3in,height=2in]{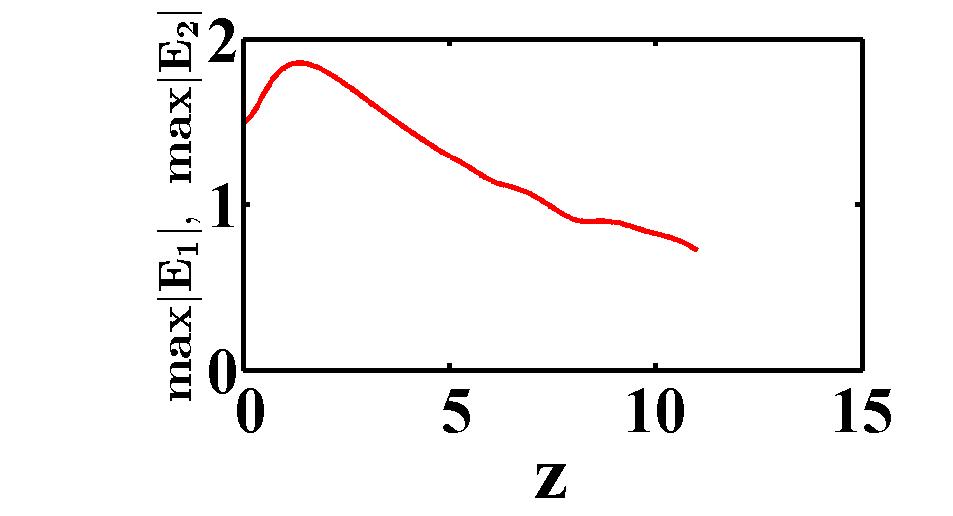}}}
      \subfigure[]{\scalebox{0.5}
{\includegraphics[width=3in,height=2in]{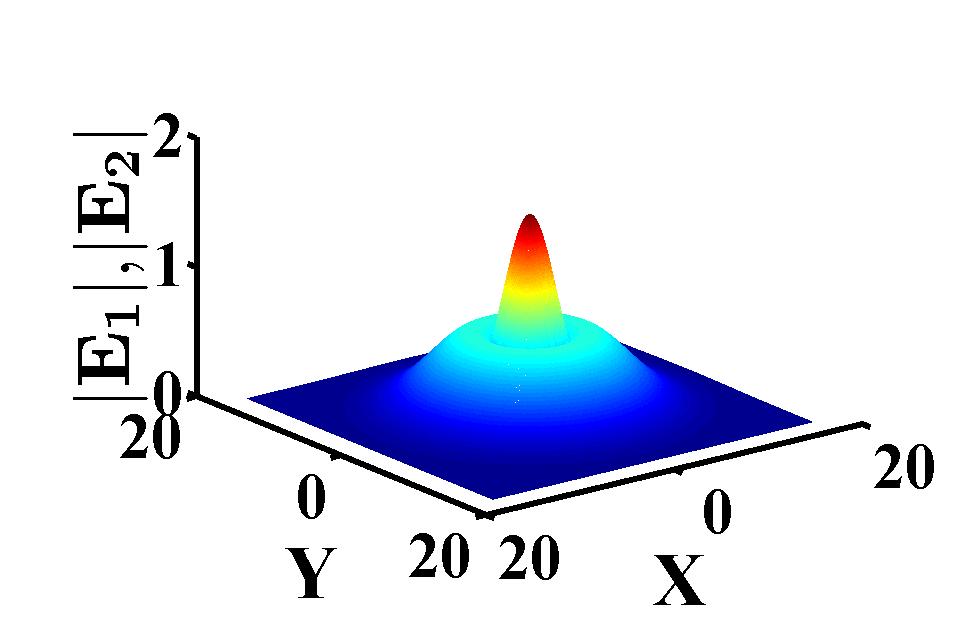}}}
      \subfigure[]{\scalebox{0.5}{\includegraphics[width=3in,height=2in]{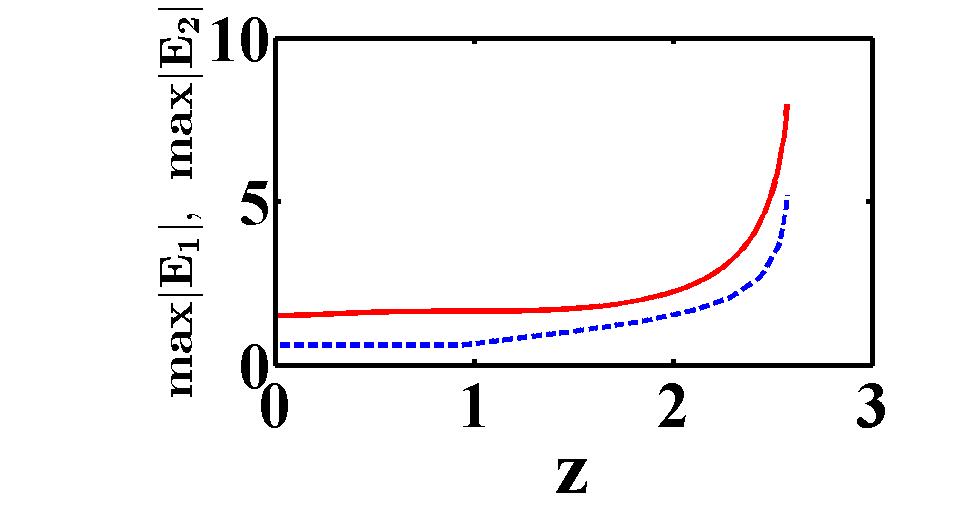}}} 
      \subfigure[]{\scalebox{0.5}{\includegraphics[width=3in,height=2in]{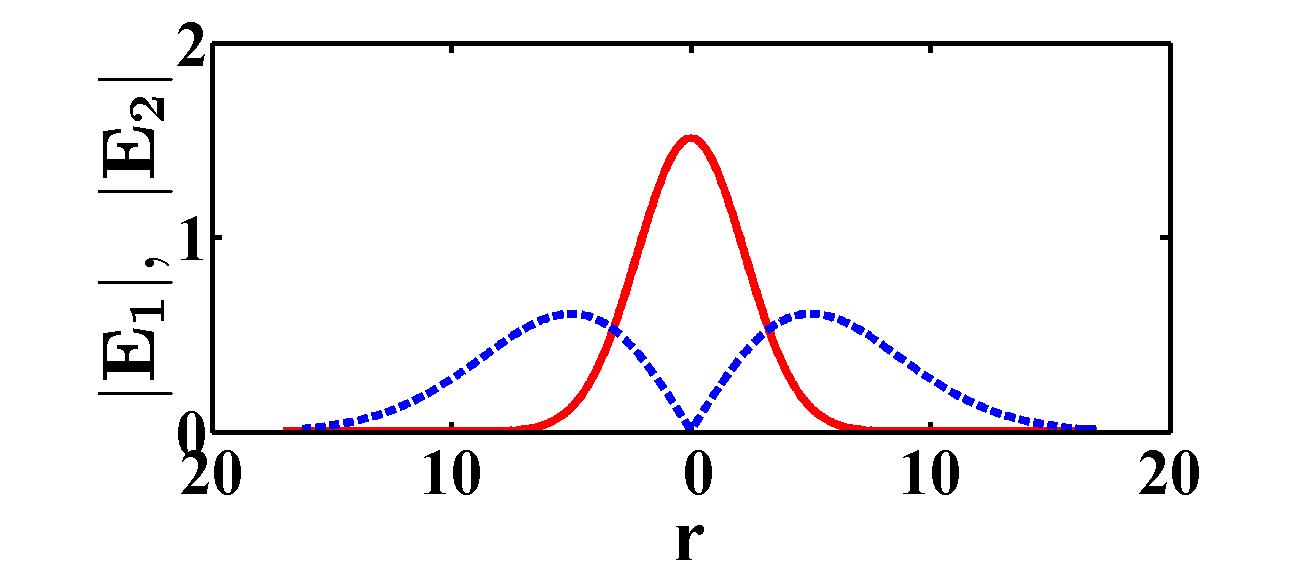}}} 
      \subfigure[]{\scalebox{0.5}{\includegraphics[width=3in,height=2in]{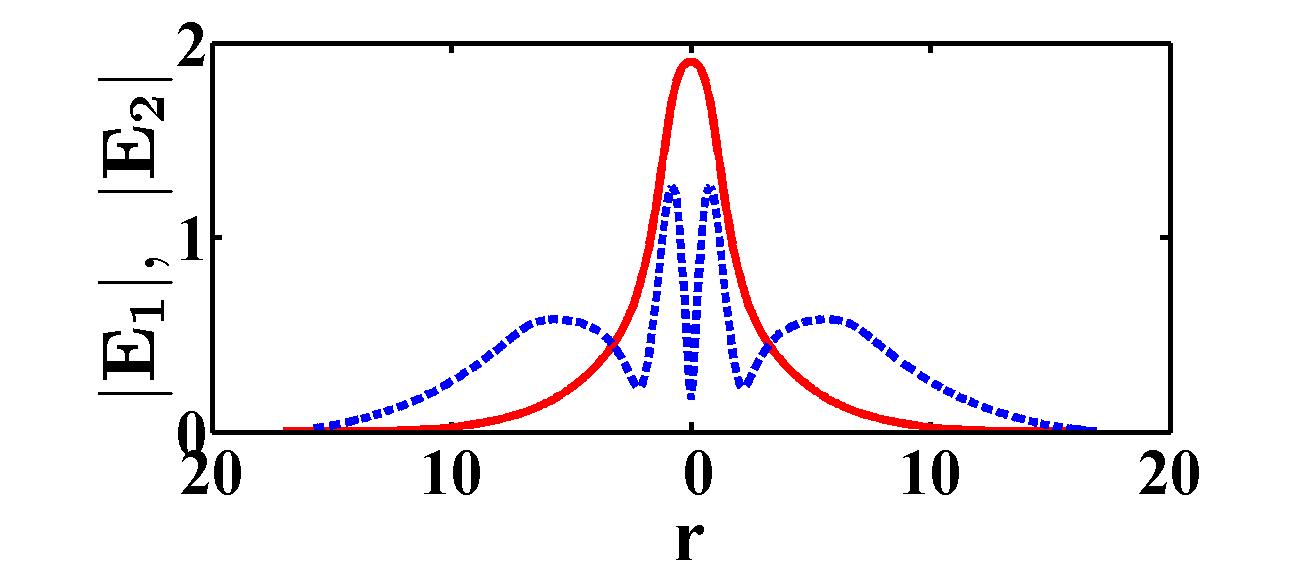}}} 
      \subfigure[]{\scalebox{0.5}
{\includegraphics[width=3in,height=2in]{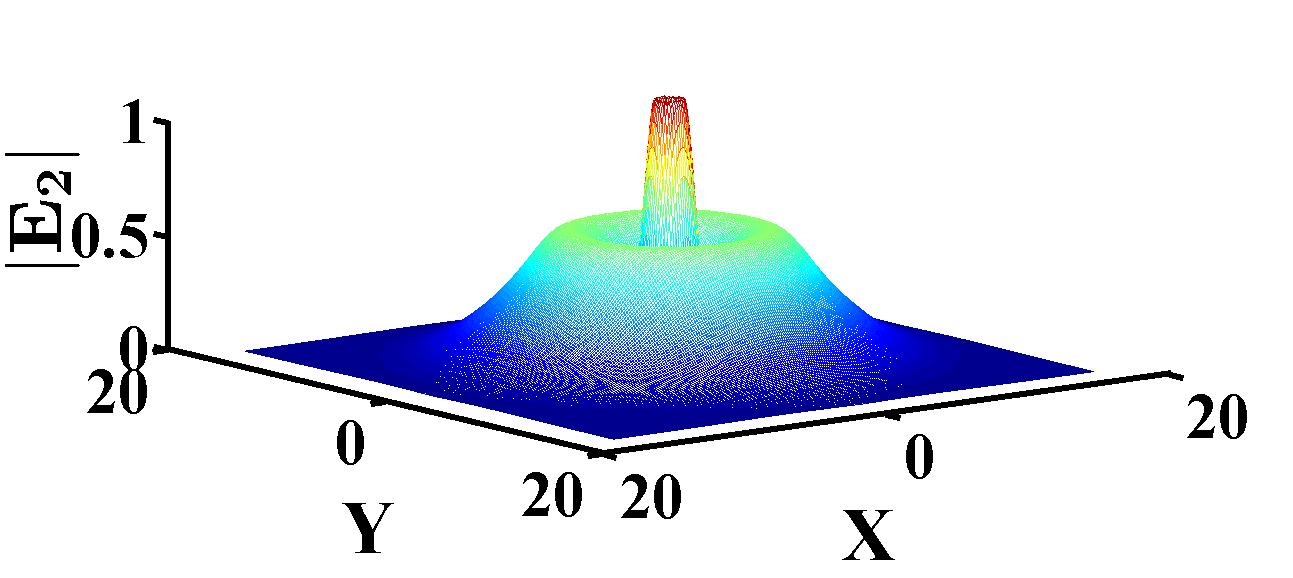}}}\quad
    \caption{Simulations parameters: $A_1=1.51, A_2=0.2, \omega_1= 3.16, \omega_2= 7.07$. (a) Initial conditions: $|E_1(z=0)|=A_1e^{-r^2/\omega_1^2}$ with $|E_2(z=0)|=0$, (b) Diffraction of the gaussian beam $E_1$, (c) Initial conditions: $|E_1(z=0)|=A_1e^{-r^2/\omega_1^2}, |E_2(z=0)|=A_2r e^{-r^2/\omega_2^2}$, (d) Collapse of the gaussian beam supported by the $m=1$ vortex beam with combined power above total power found by equations (4,5). Radial profiles of $|E_1|,|E_2|$ at z=0 (e), at z=0.7 (f), (g) Deformation of the vortex beam $E_2$ at z=0.7}
    \label{fig:m_0}
  \end{center}
\end{figure}

The case shown in figure 3 proves one can control collapse by a proper insertion of a co-propagating vortex. In Figure 3a, $E_2=0$ and a single Gaussian beam $E_1$ with power below critical, as expected diffracts during propagation Figure (3b). Notice that when $E_2=0$, the behavior of $E_1$ is governed by the 2d-NLS equation. The power of $E_1$ in this case is less then that of Townes soliton. If instead, we add a vortex mode $E_2(z=0)$, so that the total power exceeds that of the combined $(m_1,m_2)=(1,0)$, the combined field evolves to into the $\beta\rightarrow\infty$  fundamental/vortex mode. Since the power is above critical, it is the collapse ($\beta\rightarrow\infty$) mode that is achieved. Individually, during propagation  the vortex beam undergoes deformation where the topological charge seemingly changes. The outcome shown in figures 3 (c,d), highlights the fact that the initial topological charge of the vortex $E_2 (z=0)$, m=1 stays conserved and instead, due to the interaction in the central core, the Gaussian beam becomes the driver for the vortex central region growth. Also our calculations suggest that at the phase singularity of the vortex a collapse event develops. Even if the total power of the combined beams exceeds the critical power, collapse does not happen if the interacting region between fundamental and vortex beams is small.

The final case we present in figure 4, is for the propagation of two combined vortices. Small perturbation of one vortex, triggers the collapse of both. As in previous cases, it is the total power which exceeds a threshold value that produces a  collapse. If only one vortex is launched with a given individual power  $P_1$ or $P_2$ from the initial state shown in figures 4a,b, they would diffract in propagation. Instead, the combined state produces a collapse event which, unlike the fundamental/vortex case, maintains the vortex structure.

\begin{figure}[htbp]
  \begin{center}
    \mbox{
      \subfigure[]{\scalebox{0.6}
{\includegraphics[width=3in,height=2in]{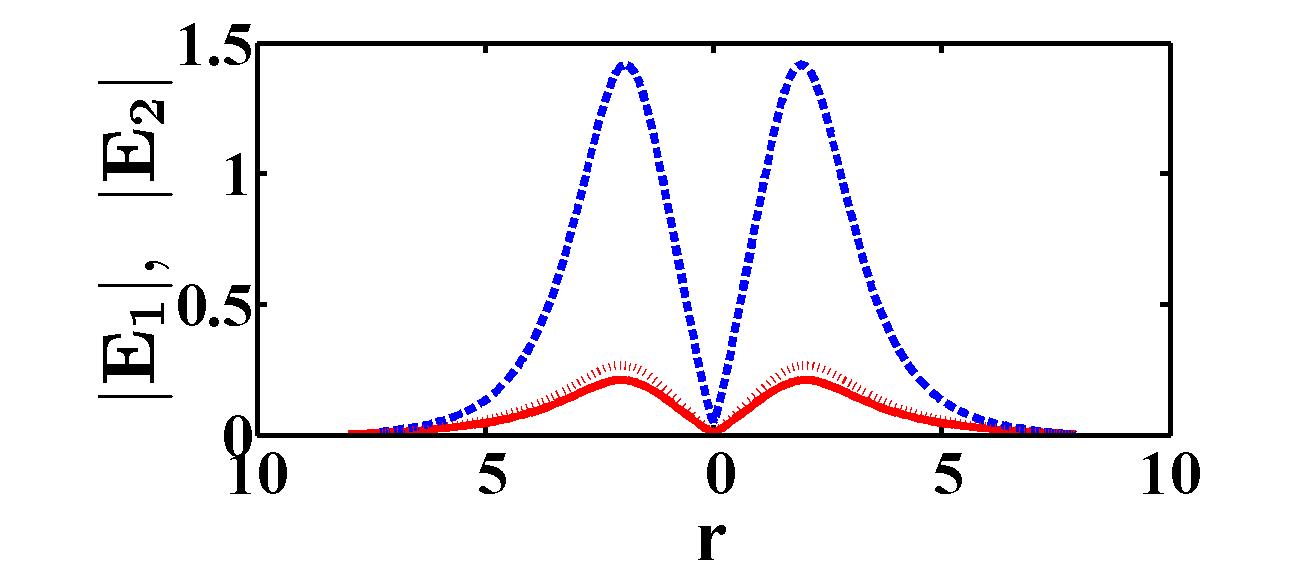}}}} \quad
      \subfigure[]{\scalebox{0.5}
{\includegraphics[width=3in,height=2in]{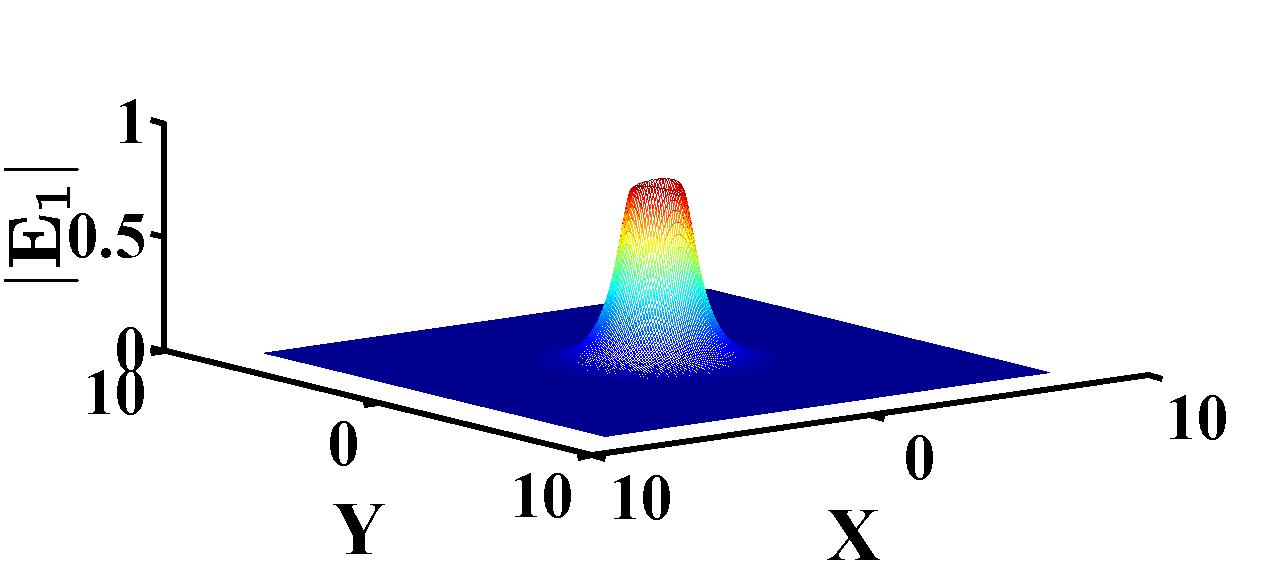}}}
      \subfigure[]{\scalebox{0.5}
{\includegraphics[width=3in,height=2in]{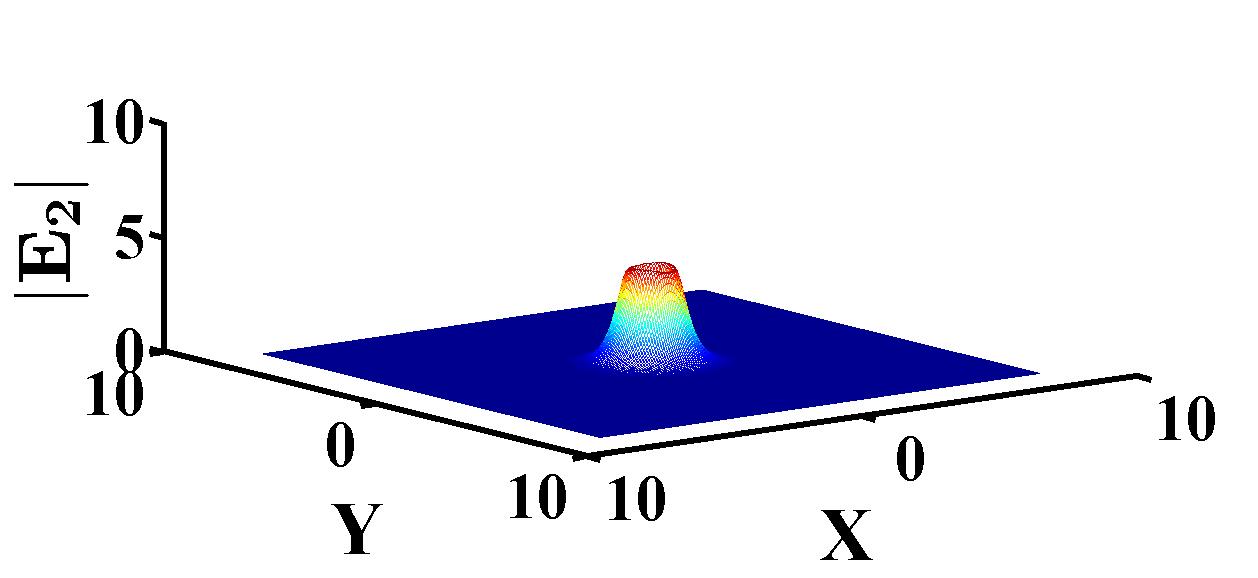}}}
    \caption{(a) Steady states ${\cal E}_1$ with power $P_1=0.2374$ (solid red line), and ${\cal E}_1$ with $P_2=7.438$ (dashed blue line). Initial conditions: $|E_1(z=0)|=1.17{\cal E}_{1}$, with power $ P_{1in}=1.3P_1$ (dotted red line), and $|E_2(z=0)|={\cal E}_{2}, (P_{2in}=P_2)$. (b,c) Collapsing (1,1)-vortex beams}
    \label{fig:m_0}
  \end{center}
\end{figure}

To conclude, we investigated the time-independent model for co-propagating beams with different frequencies. Our simulations strongly suggest that as in the 2d NLSE there is a critical {\bf combined} power together with the ratio of individual powers that determines simultaneous collapse events  even in cases where an individual component is below critical as determined by the Townes mode. The  dynamical evolution is strongly dependent on the initial state. For example near the point of collapse, the beams  converge to the ground state solution of (1,2) in the collapse ($\beta\rightarrow\infty$)limit. Depending of the initial power configuration each beam tends to collapse at different rates. We also encountered novel collapse events for configurations with topological charges beyond $m_1=m_2=0$ case. 

\section*{Acknowledgements}
This work was supported by the Army Research Office, under the MURI grant W911NF-11-0297 and partially by the US Department of Energy grant number DE-SC0011446.

\bibliography{ad,en,oz}

\end{document}